\documentclass[12pt]{article}
\usepackage{epsf}
\usepackage{alltt}
\usepackage{epsfig}
\def\largelinestretch{\renewcommand{\baselinestretch}{1.1}}
\largelinestretch\small\normalsize

\title{
\vspace*{-15mm}
{\normalsize \begin{tabular}[t]{ll}
DESY   & \hspace{-0.5em}04--105\\
June      & \hspace{-0.5em}2004\end{tabular}
  \hfill { } \\}
\vspace*{15mm}
{\bf  Bremsstrahlung in Leptonic Onia Decays:
Effects on Mass Spectra}
}

\author{
       {\it Alexander Spiridonov}\footnote{Permanent address: 
Institute of Theoretical and Experimental Physics, 117259 Moscow, Russia}\\
        DESY Zeuthen }
\date{}

\begin{document}
\maketitle
\begin{abstract}
In this note we present a study of the radiative tails in the
invariant mass spectra of muon or electron pairs from 
$J/\psi, \psi(2S)$ and $\Upsilon(1S)$ decays which is due to an
additional emission of a photon.
An analytic formula for dilepton mass spectra in radiative
decays is derived
and a Monte Carlo simulation for realistic detector
conditions is used to study effects on the spectra. 
A rather simple parameterisation is given,
suitable for the treatment of real data.
\end{abstract}

\newpage
 \section{Introduction}
The prediction from the Standard Model for radiative decays
of $Z$ bosons into lepton pairs was discussed by
Fleischer and Jegerlehner~\cite{jeger} and 
analytic results for the photon emission valid 
for arbitrary cuts on photon energies and angles
have been presented. 
The lowest order decay width $\Gamma_0(Z\rightarrow l^+l^-)$ and
contributions from radiative corrections were factorised in the
presented
formulae. We used the same formulae with appropriate mass parameters
to estimate the radiative effects for
heavy vector mesons decaying into lepton pairs.

Radiative effects in the context of the bottomonia spectroscopy were
discussed in details by Buchm\"uller and Cooper~\cite{cooper}
with a clear recommendation (p.427) that the experimental
definition of $\Upsilon\rightarrow e^+e^-$ should include all
$\Upsilon\rightarrow e^+e^-\gamma$ decays in order to avoid
sensitivity to a cut in the photon energy.  
Certainly that should be valid also for charmonia decays and
for $\mu^+\mu^-$ and $\mu^+\mu^-\gamma$ final states,
respectively. There are no principal differences between these
cases, besides the level of the radiative effects.

The first observation~\cite{e760} of the radiative 
decay $J/\psi\rightarrow e^+e^- \gamma$,
where the branching ratio 
$B (J/\psi\rightarrow e^+e^- \gamma, E_\gamma > $100 MeV)
divided by the branching ratio $B (J/\psi\rightarrow e^+e^-)$ was
measured as $0.147 \pm 0.022$, is in good agreement with the expectation.
The theoretical prediction based on results~\cite{jeger} 
for the corresponding decay into muons is smaller
by a factor about 3 only, i.e. also rather large. Radiative
tails could be statistically significant in inclusive mass spectra
of lepton pairs from decays of $J/\psi$ in measurements
with high statistics and rather good mass resolution like
in the HERA-B experiment~\cite{herab}.     

We derive an analytic formula for the dilepton invariant mass spectra 
in radiative decays. In practice, a convolution
of such ideal spectrum with detector mass resolution should
by taken into account. We investigated the smearing effects
by Monte Carlo simulation taking into account 
realistic detector conditions
for the $J/\psi, \psi(2S)$ and $\Upsilon(1S)$ decays
into $\mu^+\mu^-\gamma$ or $e^+e^-\gamma$.  
A simple parameterisation was found to describe simulated dilepton
mass spectra for different mass resolutions. 
The parameterisation fitted rather well the simulated mass spectra
and was appropriate for fitting of real dimuon spectra 
from $J/\psi$ decays. 
 
 \section{Photon Emission in Decays into Lepton Pairs}
In~\cite{jeger} radiative decays of $Z$ bosons were
considered and analytic results 
for arbitrary
cuts on the photon energy and on the angles between photon and leptons
were presented. The authors 
included the complete one-loop electroweak corrections and 
multi--soft--photon effects.
 The photons (real or virtual)
could be emitted and absorbed only by the final state leptons. 
The effect of the initial state 
entered only via the phase space boundaries given by its mass,
therefore only the mass of
the decaying particle entered into the expressions for radiative corrections
and the lowest order decay width $\Gamma_0(Z\rightarrow l^+l^-)$
was used as a normalisation factor. 
We used the appropriate formulas from~~\cite{jeger}
to study radiative decays of heavy vector mesons into lepton pairs.

Let us consider the decay $X\rightarrow l^+l^-$ of
a heavy vector state with mass $M$ into a pair of leptons with masses $m_l$.  
The lowest order decay width is
\begin{equation} 
\Gamma_0 = \Gamma_{0}(X\rightarrow l^+l^-).
\label{g0}         
\end{equation}           
The bremsstrahlung process
\begin{equation}    
 X(p_0) \rightarrow l^-(p_1) + l^+(p_2) + \gamma(k) 
\label{raddecay}           
\end{equation}             
is distributed in the phase space as~\cite{jeger}:
\[  \frac {1}{\Gamma_0}~~\frac {d^2\Gamma(X\rightarrow l^+l^-\gamma)}
{d\varsigma d\tau}
= P(\varsigma,\tau) \]
\begin{equation}   
= \frac {\alpha}{2 \pi} \left[ \left( 
 \frac {1 + \varsigma^2}{1 - \varsigma}\right) 
\left( \frac {1}{\tau} + \frac {1}{1 - \varsigma - \tau} \right) 
 - \frac {a}{2} \left( \frac {1}{\tau^2} 
+ \frac {1}{(1 - \varsigma - \tau)^2} \right) - 2~~\right],   
\label{gtot}          
\end{equation}            
where $\alpha$ is the fine structure constant, 
$\varsigma = (p_1 + p_2)^2 / M^2$,
$\tau = (p_0 - p_1)^2 / M^2$ and $a = 4{m_l}^2/M^2$,
$a \le \varsigma \le 1$. 

In the rest frame of the decaying particle we have 
$\varsigma = 1 - 2E_{\gamma}/M, \tau = 1 - 2E_1/M$ and
$1 - \varsigma - \tau = 1 - 2E_2/M$,  
and in the center of mass system of the lepton pair:
\[ \tau = 1 - \varsigma - \frac {1}{2} (1 - \varsigma) 
(1 - r\cos \theta_1)\,\,\,\, \rm{or}
\,\,\,\,  \tau = \frac {1}{2} (1 - \varsigma)(1 - r~\cos \theta_2) \]
with $r = \sqrt{1 - a/\varsigma}$
and $\theta_i$ the angle between the
photon and lepton with momentum $\vec{p_i}$, 
obviously $\theta_1 = \pi - \theta_2$.
Variations of $\theta_2$ from $\theta_2 = 0$ to $\theta_2 = \pi$
corresponds to 
\[ \tau(0) = \frac {1}{2} (1 - \varsigma)(1 - r),~~
\tau(\pi) = \frac {1}{2} (1 - \varsigma)(1 + r). \]
The parameter $\varsigma$ is related to the invariant mass of
the lepton pair:
\[ m^2 = (p_1 + p_2)^2 = \varsigma \cdot M^2.\] 
The distribution $P(\varsigma)$ can be
evaluated by integrating the distribution $P(\varsigma,\tau)$ over
the parameter $\tau$:
\[ \frac {1}{\Gamma_0}\,
\frac {d\,\Gamma(X\rightarrow l^+l^-\gamma)}{d\,\varsigma}
= P(\varsigma) = \int_{\tau(0)}^{\tau(\pi)} P(\varsigma,\tau)\,d\tau \]
\begin{equation}  
= \frac {\alpha}{\pi}\,\, 
\frac {1 + \varsigma^2}{1 - \varsigma}\,\,  
\left( \ln\frac {1 + r}{1 - r}  -  r \right),
\label{psigma}           
\end{equation}             
The latter formula agrees with the expression 
$P(\varsigma, > \delta)$ from ~\cite{jeger} in the limit $\delta = 0$
presenting the result of integration of $P(\varsigma, \tau)$ over 
the angles $\theta_1,\theta_2$ with exclusion of cones around the leptons
defined by the angle $\delta$.  

\begin{figure}[ht]
\begin{center}
\epsfxsize=0.5\textwidth
\epsfbox{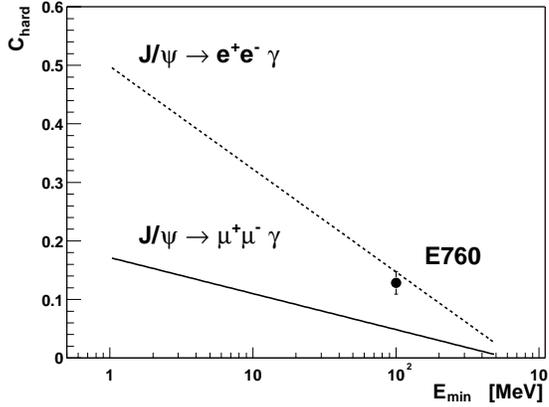}
\caption{\small Parameter $C_{hard}$ as a function
of the minimal energy $E_{min}$ of the photon in the $J/\psi$ rest frame  
for radiative decays $J/\psi \rightarrow \mu^+\mu^-\gamma$ (solid line) 
and $J/\psi \rightarrow e^+e^-\gamma$ (dashed line).
The point with errors was evaluated from the $E760$ result~\cite{e760}.} 
\label{e760}
\end{center}
\end{figure}

Photons with sufficient energy can be detected.
The fraction of decays corresponding to the emission of hard photons is
\begin{equation}   
C_{hard}(E_{min})\,\,
=\,\,\frac {1}{\Gamma_0} \cdot 
\Gamma(X\rightarrow l^+l^-\gamma, E_\gamma > E_{min})\,
=\,\,\int_a^{1 - 2E_{min}/M}P(\varsigma)\,d\varsigma,
\label{chardint}            
\end{equation}              
where $E_{min}$ is the minimal photon energy in the rest frame of the decaying
particle. 
The result of integration for $E_{min} \ll M/2$, neglecting terms
of order $O(a\alpha)$, is known~\cite{jeger}: 
\begin{equation}   
C_{hard}(E_{min})\,=\,\frac{\alpha}{2\pi}\,
\left[ 4\ln \frac{M}{2E_{min}} \left( \ln \frac{M^2}{m_l^2} - 1 \right)
- 3 \ln \frac{M^2}{m_l^2} 
- \frac {2}{3} \pi^2 + \frac {11}{2} \right].
\label{chard}           
\end{equation}             
The measurement of $J/\psi\rightarrow e^+e^- \gamma$~~\cite{e760}
is in good agreement with the calculation $C_{hard}$ in fig.~\ref{e760}.

\begin{figure}[ht]
\begin{center}
\unitlength1cm
\begin{picture}(15.5,6)
\put( -.50,0.0){\epsfig
{file=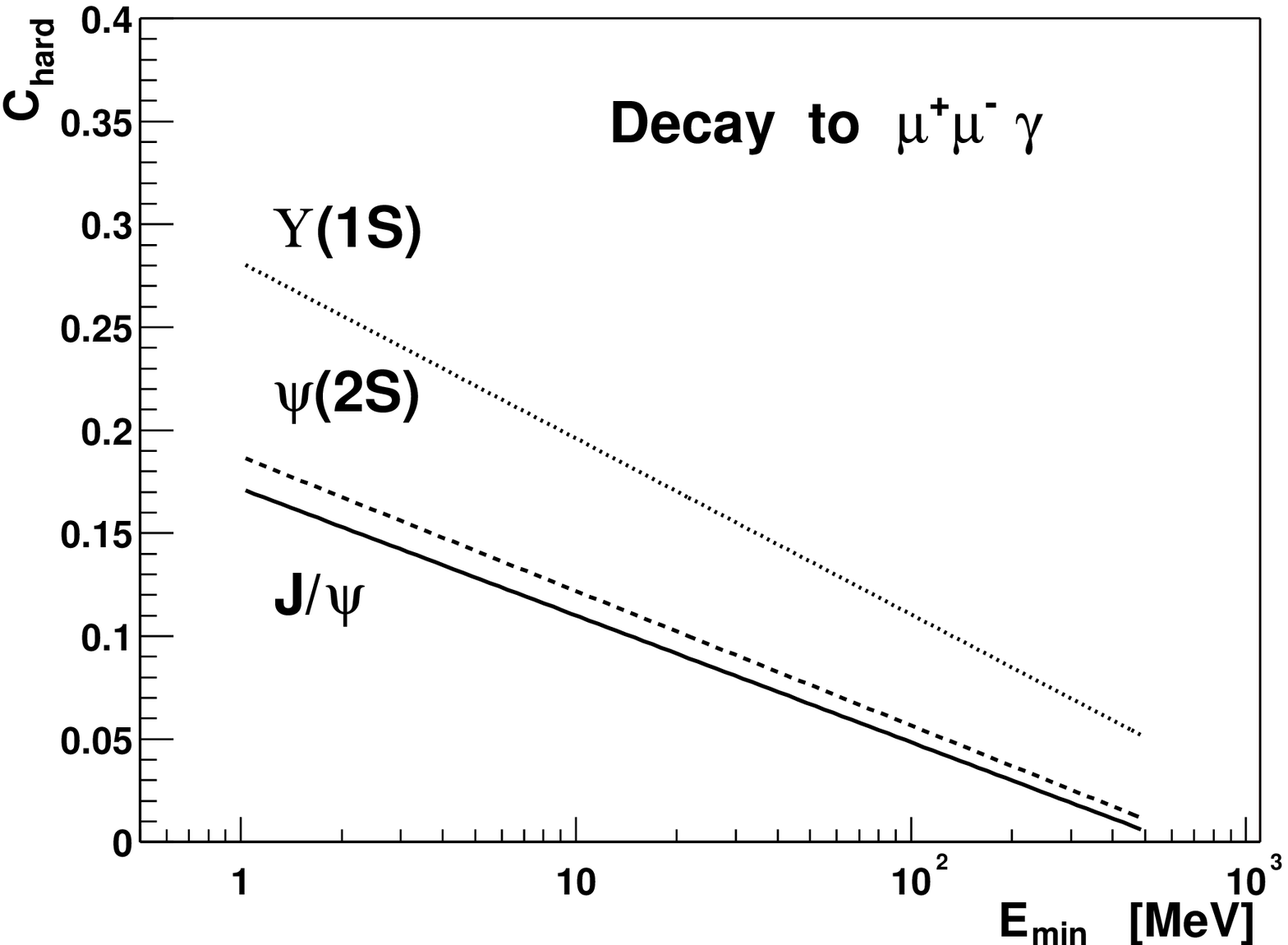,width=8.9cm}}
\put(7.0,3.2){\makebox(0,0)[t]{\huge a)}}
\put(7.7,0.0){\epsfig
{file=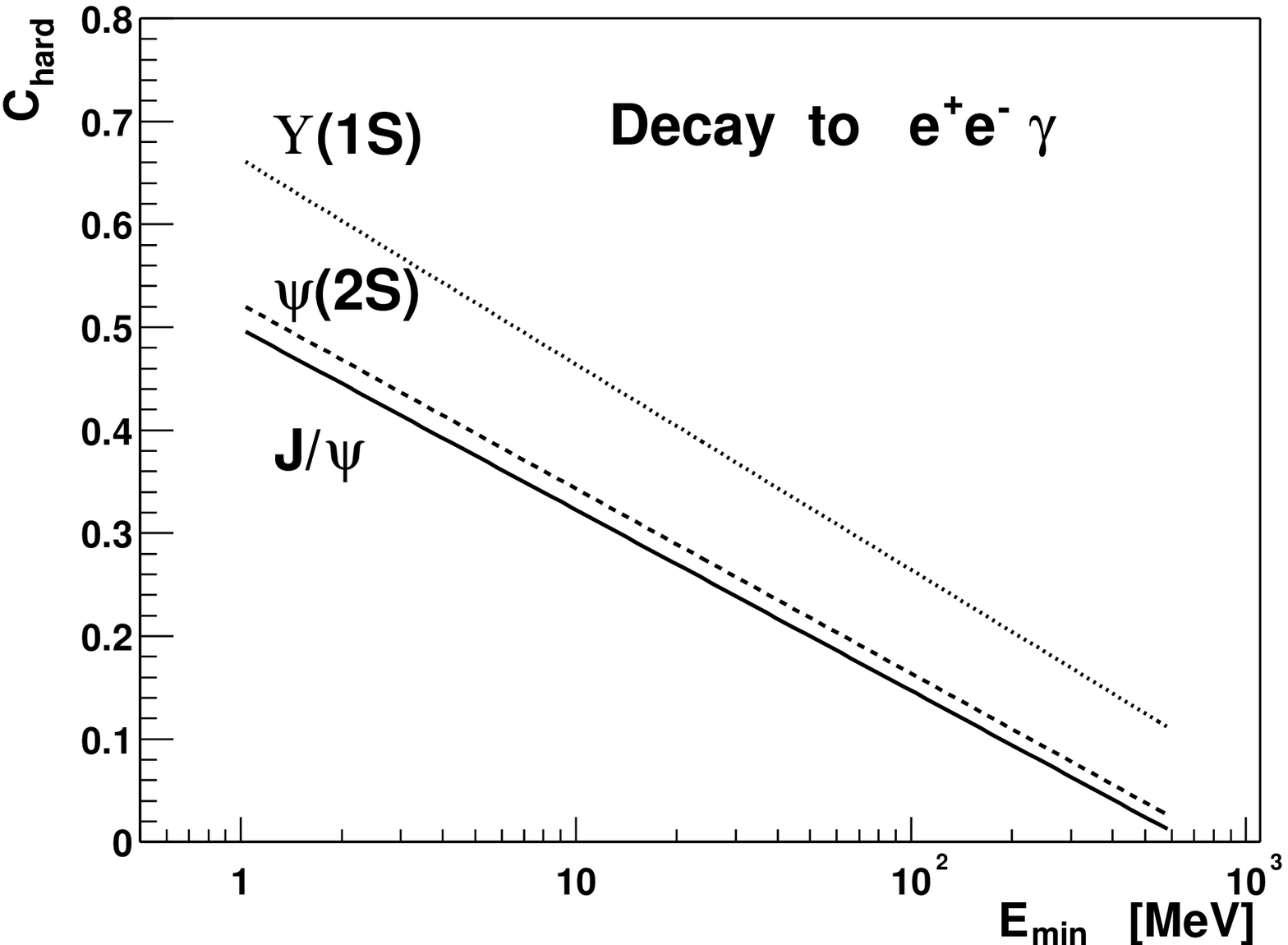,width=8.9cm}}
\put(15.0,3.2){\makebox(0,0)[t]{\huge b)}}
\end{picture}
\caption{\small 
The parameter $C_{hard}$ as a function
of the minimal energy $E_{min}$ of the photon in the rest frame  
of $J/\psi$ (solid line),
$\psi(2S)$ (dashed line), and $\Upsilon(1S)$ (dotted line)
decaying into $\mu^+\mu^-\gamma$ (a) or $e^+e^-\gamma$ (b).}
\label{allmue}
\end{center}
\end{figure}

The lepton mass entered in eq.~\ref{chard} 
in the first (leading) term as
$\ln(M^2/m_l^2)$ which is smaller only by a factor 2.6
for muons compared to electrons. 
The effect of hard photon emission
is expected to be rather large also for decays 
into $\mu^+\mu^-\,\gamma$ as shown in fig.~\ref{e760}.
The effects are larger for heavier decaying particles as seen
in fig.~\ref{allmue} comparing $J/\psi,~\psi(2S)$ and $\Upsilon(1S)$,
respectively.

\begin{figure}[ht]
\begin{center}
\epsfxsize=0.5\textwidth
\epsfbox{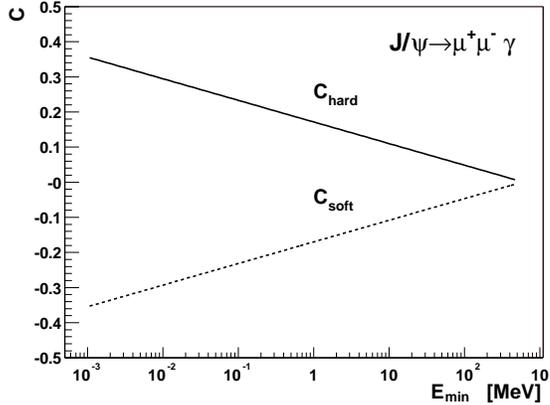}
\caption{\small The parameters $C_{hard}$ (solid line) and
$C_{soft}$ (dashed line) for the radiative decay 
$J/\psi \rightarrow \mu^+\mu^-\gamma$ as functions
of the minimal photon energy $E_{min}$ in the rest frame of the $J/\psi$.}  
\label{chcs}
\end{center}
\end{figure}

In order to get the physical rate, we have to take into account 
contributions
from virtual photons (emitted and absorbed by leptons) and 
emitted soft photons $(E_\gamma < E_{min})$. 
The sum of these contributions is known~\cite{jeger}: 
\begin{equation}   
C_{soft}(E_{min})\,=\,\frac{\alpha}{2\pi}\,
\left[ -4\ln \frac{M}{2E_{min}} \left( \ln \frac{M^2}{m_l^2} - 1 \right)
+ 3 \ln \frac{M^2}{m_l^2} 
+ \frac {2}{3} \pi^2 - 4 \right].
\label{csoft}           
\end{equation}             
The total width taking into account $l^+l^-,\,l^+l^-\,\gamma$ configurations
is 
\begin{equation}
\Gamma_{all}(X\rightarrow l^+l^-, \, l^+l^-\gamma) =
(1 + C_{soft} + C_{hard} ) \, \cdot \Gamma_0 (X\rightarrow l^+l^-).    
\label{gtotal}            
\end{equation}              
The term $C_{soft}$ is negative as seen in fig.~\ref{chcs}.
The leading terms in $C_{soft}$ and $C_{hard}$ have singular
behavior for $E_{min} \rightarrow 0$, but  
in the sum $C_{soft} + C_{hard}$ 
they exactly cancel, as it should be:    
\begin{equation} 
C_{soft} + C_{hard} = \frac {\alpha}{2\pi} \, \frac {3}{2}.
\label{cschsum}             
\end{equation}               

When an experiment measures all $l^+l^-,\,l^+l^-\,\gamma$ 
configurations
independent of the energy of emitted photons,
the effects
of radiative corrections appear as a small increase of
the decay width
\begin{equation}  
\frac {\Gamma_{all} - \Gamma_0}{\Gamma_0}
= \frac {\alpha}{2\pi} \, \frac {3}{2} \approx 0.00174 .
\label{dgtotal}              
\end{equation}                
However, cuts on the energy of the emitted photon can have
much stronger effect. 
Let us assume that we are 
selecting lepton pairs inclusively 
and apply a cut on their invariant mass
\[      | m - M |\,\, <\,\,\Delta. \]
The radiative decays with photon energies in the rest frame of the decay 
\begin{equation}
E_\gamma > E_\Delta = \Delta\,\left(1 - \frac {\Delta}{2 M} \right)   
\label{edelta}               
\end{equation}                 
will be rejected and the number of selected decays will be
reduced by the factor 
\begin{equation}
\frac {N(l^+l^-,\,| m - M |\, <\,\Delta)}{N(l^+l^-)}\,\,=\,\,
\frac {\left( 1 + C_{soft}(E_\Delta)\right)}
{\left( 1 + \frac {\alpha}{2\pi} \, \frac {3}{2} \right)}.   
\label{fractdelta}               
\end{equation}                 
For inclusive registration of $\mu^+\mu^-$ pairs
from  $J/\psi$  decays and the rather soft  cut  $\Delta\,=\,200$\,MeV,
about $3\%$ of the decays will be rejected.
In general, the dilepton mass $m$ is shifted by photon emission 
\begin{equation}   
m = \sqrt{M (M - 2 E_\gamma)} \approx M - E_\gamma~~~(E_\gamma \ll M).
\label{dmass}               
\end{equation}                 
\begin{figure}[h]
\begin{center}
\epsfxsize=0.5\textwidth
\epsfbox{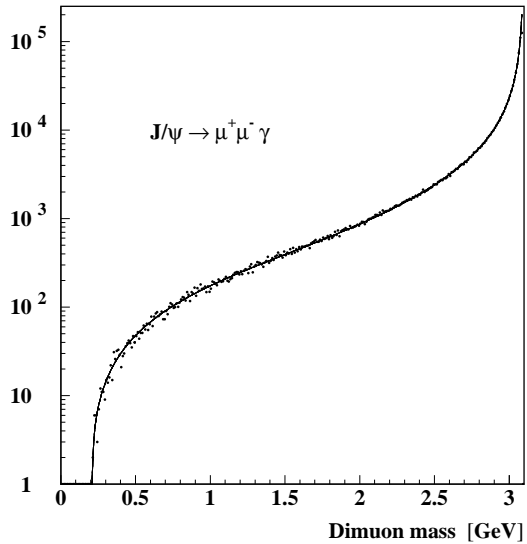}
\caption{\small 
The dimuon mass spectrum defined by eq.~\ref{pmass} (solid line)
and results of Monte Carlo simulation (points) 
of the decay $J/\psi \rightarrow \mu^+\mu^-\gamma$ for an ideal detector
(without smearing, $4\pi$ acceptance) for $E_{min}$ = 10 MeV.} 
\label{dndmideal}
\end{center}
\end{figure}
Hard photon emission causes a tail towards lower masses in the 
dilepton mass spectrum. 
The distribution $P(m)$ of the dilepton mass in the radiative 
decay (\ref{raddecay}) 
we find from eq.~(\ref{psigma}) as
\begin{equation}  
\frac {1}{\Gamma_0}\, 
\frac {d\,\Gamma(X\rightarrow l^+l^-\gamma)}{d\,m} = P(m) 
= \frac {\alpha}{\pi} \frac {2 m}{(M^2 - m^2)}
 \left(1 + \frac {m^4}{M^4}\right) 
\left( \ln\frac {1 + r}{1 - r}  -  r \right),
\label{pmass}           
\end{equation}             
where $r = \sqrt{1 - 4{m_l}^2 / m^2}$ 
is also a function of $m$.
The singular behavior of the spectrum for $E_\gamma \rightarrow 0$
($m \rightarrow M$) 
is clearly seen in fig.~\ref{dndmideal}.

In an experiment which measures the 
dilepton mass spectrum inclusively 
a separation of $l^+l^-$ and $l^+l^-\gamma$ final states 
is not possible on an event by event basis.
For very low photon energies that is not possible even 
on a statistical basis, because both decays contribute
to the peak around the mass of the decaying particle.
The measured spectrum
can be described by two components. 
The main component corresponds
to decays into $l^+l^-$ and $l^+l^-\gamma$ ($E_\gamma \le E_{min}$)
contributing
to a ``monochromatic'' line ($m = M$) in dilepton mass spectrum 
smeared by detector resolutions.
An additional component is related to   
the $l^+l^-\gamma$ $(E_\gamma > E_{min})$ final states 
producing the spectrum (\ref{pmass}) also affected
by detector resolutions. 
The fraction of events in the latter component is
\begin{equation} 
W_{hard} = \frac {C_{hard}(E_{min})}
{1 + \frac {\alpha}{2\pi} \, \frac {3}{2}}.  
\label{whard}   
\end{equation}   

 \section{Smearing Effects due to Detector Resolutions}
\begin{figure}[ht]
\begin{center}
\epsfxsize=0.5\textwidth
\epsfbox{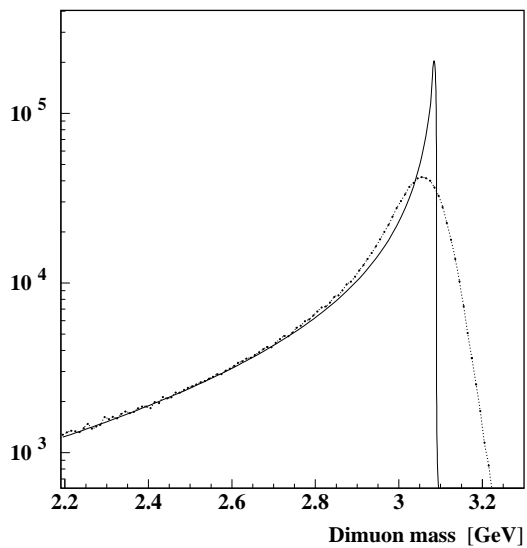}
\caption{\small Dimuon mass spectra in the radiative decay 
$J/\psi \rightarrow \mu^+\mu^-\gamma$ for $E_{min} = 10$ MeV (solid line) 
smeared 
with mass resolution of about 42 MeV (points connected by dashed line).} 
\label{smearing}
\end{center}
\end{figure}

After smearing due to limited detector resolutions, 
the shape of the dilepton mass spectrum from
radiative decay (\ref{pmass}) changes in particular
for low energy  photons ($m\approx M)$ as seen  
in fig.~\ref{smearing}.
The spectrum (\ref{pmass})
convoluted with the Gaussian could not be expressed 
in an analytic form and in addition, experimental cuts
complicate the situation even more.

Our goal was to derive a function to be used in fitting the spectra. 
We investigated radiative decays by Monte Carlo 
simulation for 
real experimental conditions
and propose a rather simple parameterisation for smeared dilepton 
mass spectra. 
We used the Monte Carlo generator for simulation of the kinematics of the 
decay (\ref{raddecay}) developed by Lohse~\cite{generator}
and based on the results in~\cite{jeger}. 
The generated spectrum of the dilepton invariant mass was in
good agreement with the analytic expression (\ref{pmass}) as shown
in fig.~\ref{dndmideal}. 
To study the real experimental conditions we 
simulated radiative decay within the framework of the HERA-B
detector. Protons with an energy of 920 GeV colliding
with a nuclear target directly produced the $J/\psi$
(or $\psi(2S),\Upsilon(1S)$). 
The distributions of the transverse momentum $p_T$
and the Feynman scaling variable $x_F$ were generated according
to the model described in~\cite{pythia}. The particle decayed into 
$\mu^+\mu^-(e^+e^-)$  or $\mu^+\mu^-\gamma(e^+e^-\gamma)$ with
a probability determined by the cut parameter $E_{min}$, the minimal energy 
of the photon in the rest frame of the decay.

The HERA-B detector is a forward spectrometer with a tracking
system including the vertex detector placed in front of the magnet
and main tracker system behind the magnet. The electro-magnetic
calorimeter and muon system located behind the main tracker are 
used for particle identification and triggering. 
The main tracker includes the
inner and outer trackers, 
but only the latter was taken into account since only this part was used
in the dilepton trigger.
Events where the both leptons passed 
through the tracking and particle identification detectors were considered.
The trigger also imposed cuts on leptons, e.g. only muons with
energy higher then 6 GeV were selected.

The momentum resolution for muons was simulated according to 
the parameterisation~\cite{momres}
\begin{equation}
\frac{\sigma(p)}{p}(\%) = 
0.94 \oplus 9.55\cdot10^{-3}\cdot p \oplus 0.484
\cdot p^{0.263},  
\label{pardppc}
\end{equation}
where $p$ is in GeV.
We were interested to isolate effects induced by the internal
bremsstrahlung in decays and therefore did not simulate 
bremsstrahlung photons from electrons on the path through the detector
material. The simulated momentum resolution for electrons was  
the same as for muons.   
To study dilepton spectra for various resolutions we multiplied
eq.~\ref{pardppc} by fudge factors from 0.7
up to 2.5. That corresponded to the variation of mass resolution at the
$J/\psi$ peak from 23 MeV up to 81 MeV, respectively. 

The decays corresponding to $E_\gamma \le E_{min}$ were simulated
as $l^+l^-$ final states
smeared by detector resolutions.
The remaining part of the spectrum consisted of simulated
$l^+l^-\gamma$ configurations $(E_\gamma > E_{min})$, also affected
by detector resolutions. Our aim was to find 
a simple analytic approximation for the latter part of the 
simulated spectrum
with the additional condition 
that the mass resolution can also be treated as unknown and included
in the fit parameters.
We solved this problem in three steps:
\begin{itemize}
\item
Find the value $E_{min}$ appropriate
for the simulation;
\item
Define a parameterisation for
the smeared dilepton mass spectra in the radiative decays; 
\item
Determine the corresponding parameters as 
functions of the mass resolution.  
\end{itemize}
\clearpage
\begin{figure}[h]
\begin{center}
\epsfxsize=0.5\textwidth
\epsfbox{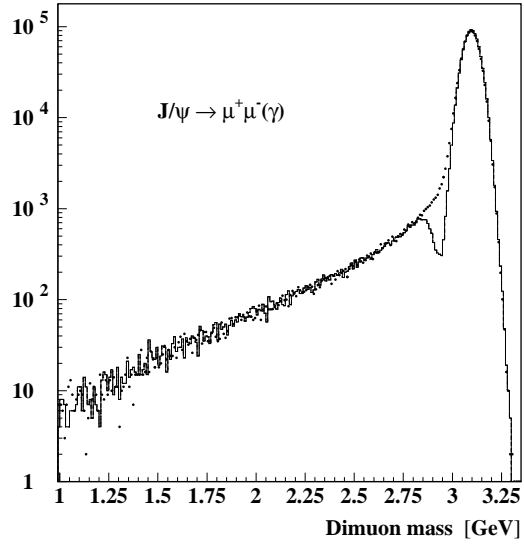}
\caption{\small
Dimuon mass spectra simulated by Monte Carlo  for the
mixture of $J/\psi \rightarrow \mu^+\mu^-$ and
$\mu^+\mu^-\gamma$ decays. The latter decay
was evaluated with the probabilities corresponding to 
$E_{min}=$10 MeV (points) and 200 MeV (histogram), respectively.
The simulated mass resolution was about 42 MeV at the $J/\psi$ peak.}
\label{dndmreal}
\end{center}
\end{figure}
\begin{figure}[h]
\begin{center}
\epsfxsize=0.5\textwidth
\epsfbox{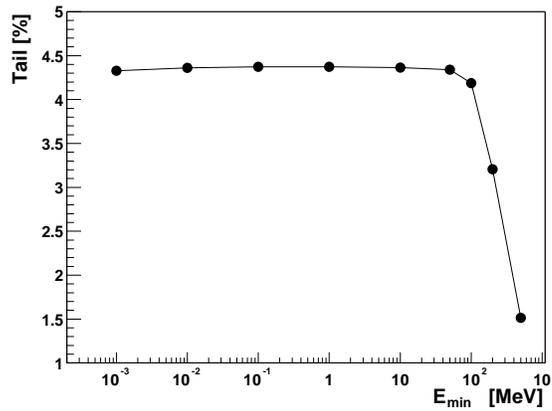}
\caption{\small
Fraction of events in the tail of dimuon mass spectra
lower then 100 MeV as the $J/\psi$ mass as a function of the cut 
parameter $E_{min}$. This fraction was 
evaluated for the mixture of $J/\psi \rightarrow \mu^+\mu^-$ and 
$\mu^+\mu^-\gamma$ decays simulated with probabilities 
corresponding to $E_{min}$.  
The mass resolution was about 42 MeV at the $J/\psi$ peak.} 
\label{tail}
\end{center}
\end{figure}
\clearpage
On the one hand, $E_{min}$ should not be extraordinarily small to
stay away from the 
singularities in eqs.(\ref{chard},\ref{csoft}). 
On the other hand, the simulation with $E_{min}$ much bigger than
the mass resolution produces unreasonable mass spectra,
as is clearly seen in fig.~\ref{dndmreal} for the 
spectrum simulated with $E_{min} =200$ MeV and a 
mass resolution of 42 MeV.   
If the photon energy is small enough compared 
to the mass resolution $\sigma$,  
it does not matter whether  $l^+l^-$ or $l^+l^-\gamma$ 
was simulated, because the invariant mass is smeared with
the resolution $\sigma \gg M - m$. The shape of the inclusive
dilepton mass spectrum becomes stable against a  
decrease of $E_{min}$.   

To make a proper choice we
estimated the fraction of events in the ``tail'' of
the dilepton mass spectra $(m  <  M(J/\psi) - 100$ MeV) 
as a function of the parameter $E_{min}$ 
for decays $J/\psi\rightarrow\mu^+\mu^-,\mu^+\mu^-\gamma$ 
simulated with the mass resolution 42 MeV.
This dependence, presented in fig.~\ref{tail}, shows a
plateau where the results are stable with respect 
to the cut parameter $E_{min}$.
We selected the value $E_{min} = 10$ MeV for further simulation.  

 \section{Parameterisation of Smeared Mass Spectra}
In this section the parameterisation of the smeared dilepton
mass spectrum from the radiative decay (\ref{raddecay}) 
with $E_{\gamma} > E_{min}$ is given.
The upper edge of the spectrum 
is the result of smearing of the sharp peak 
in the dilepton mass spectrum (\ref{pmass})
at $m \approx M$ and it looks like  
a Gaussian (see fig.~\ref{smearing}).
For the full dilepton mass range we propose the
following parameterisation: 
\begin{equation}
{\cal P}(m) = G (m | m_R, \sigma_R),~~m \ge m_S  
\label{gauss} 
\end{equation} 
and 
\begin{equation}
{\cal P}(m) = c_S \cdot G ( m | m_R, \sigma_S(m)),~~m < m_S,  
\label{gtail} 
\end{equation} 
where $G ( m | m_R, \sigma_R)$ is the Gaussian function
\[ G ( m | m_R, \sigma_R) = \frac {1}{\sqrt{2 \pi}\,\sigma_R}\,
\exp \left( - \frac {(m - m_R)^2}{2 \sigma_R^2} \right) \]
with an average $m_R$ and standard deviation $\sigma_R$,
and $G ( m | m_R, \sigma_S(m))$ that with standard deviation $\sigma_S(m)$,
respectively,
and $c_S$ is a scaling factor. For a given mass resolution,
the parameters $m_R, \sigma_R$ and $c_S$ are considered as constants, 
but $\sigma_S(m)$ as a polynomial: 
\begin{equation} 
\sigma_S(m) = 
\sum_i a_i^2\,(m_R - m)^i
\label{sigmas}  
\end{equation}  
with constants $a_i$ to be defined.
Each term in the polynomial is positive which
helps to reduce mutual correlations of parameters $a_i$
during their search. 

In the point $m = m_S$ we require continuity
of the function ${\cal P}(m)$ and its first derivative
${\cal P}^\prime(m)$. From the first condition we
define the parameter $c_S$:
\begin{equation}  
c_S = \frac {\sigma_S(m_S)}{\sigma_R}
\exp \left( \frac {(m_S - m_R)^2}{2\,\sigma_S^2(m_s)}  
- \frac {(m_S - m_R)^2}{2\,\sigma_R^2} \right).
\label{cs}  
\end{equation}  
From the second condition follows the nonlinear equation
\begin{equation}  
\left[ \frac {1}{\sigma_S(m)} \, \frac {d\,\sigma_S(m)}{d\,m}
+ \frac {m - m_R}{\sigma_S^2(m)}
- \frac {(m - m_R)^2}{\sigma_S^3(m)}\, 
\frac {d\,\sigma_S(m)}{d\,m} \right]_{m=m_s}
= \frac {m_S - m_R}{\sigma_R^2} 
\label{ms}  
\end{equation}  
to define $m_S$.
\begin{figure}[ht]
\begin{center}
\epsfxsize=0.5\textwidth
\epsfbox{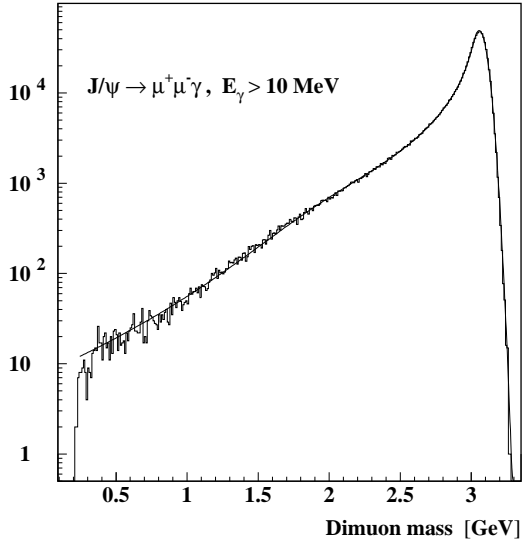}
\caption{\small
Monte Carlo results for the dimuon mass spectrum from the radiative decay
$J/\psi \rightarrow \mu^+\mu^-\gamma$ (histogram) and the 
parameterisation described in text (solid line). 
The simulation was made for 
a mass resolution of about 42 MeV
at the $J/\psi$ peak.}
\label{radtailjpsi}
\end{center}
\end{figure}
\begin{figure}[ht]
\begin{center}
\unitlength1cm
\begin{picture}(15.5,8)
\put( -.50,0.0){\epsfig
{file=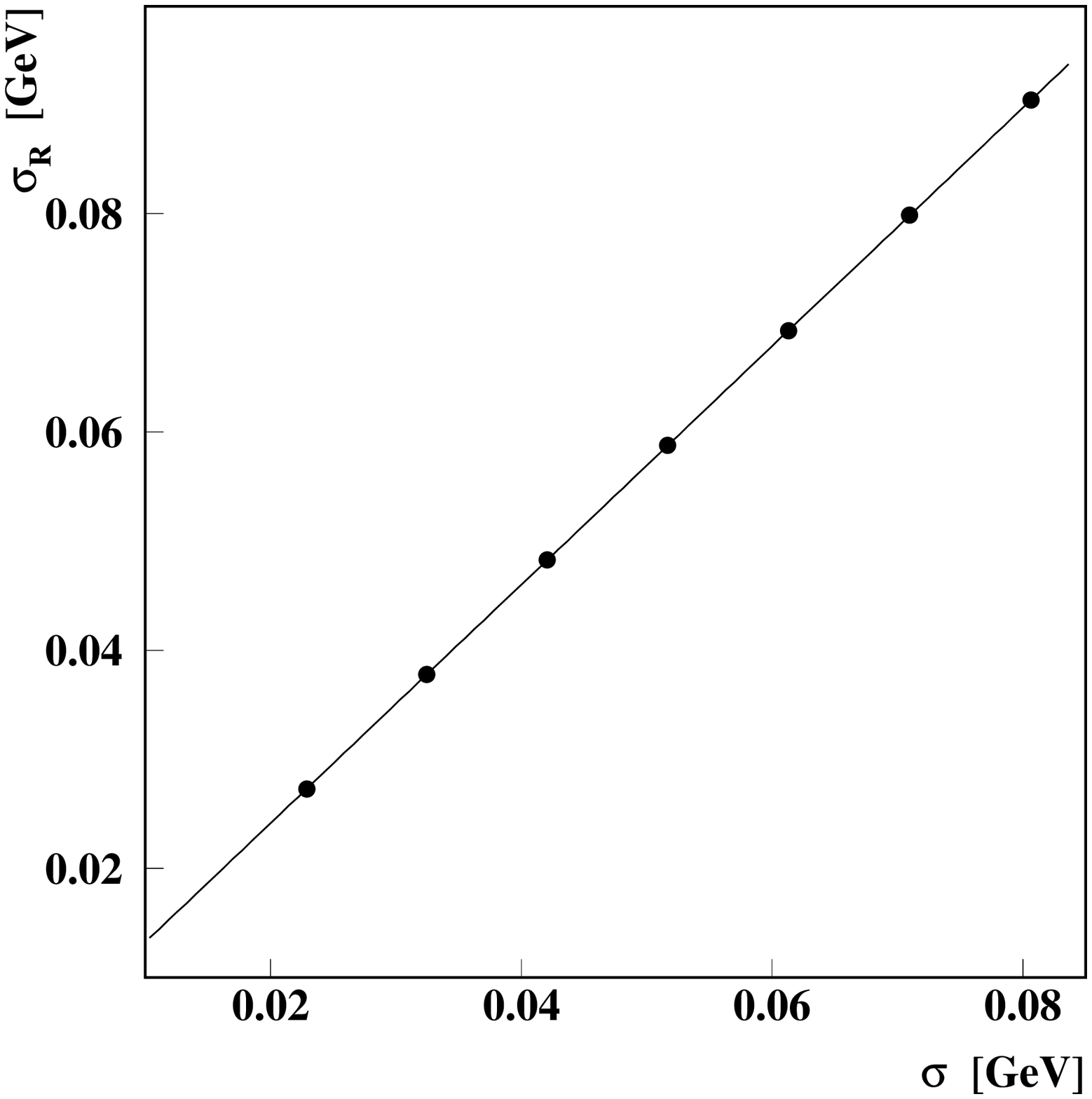,width=8.9cm}}
\put(7.0,4.2){\makebox(0,0)[t]{\huge a)}}
\put(7.7,0.0){\epsfig
{file=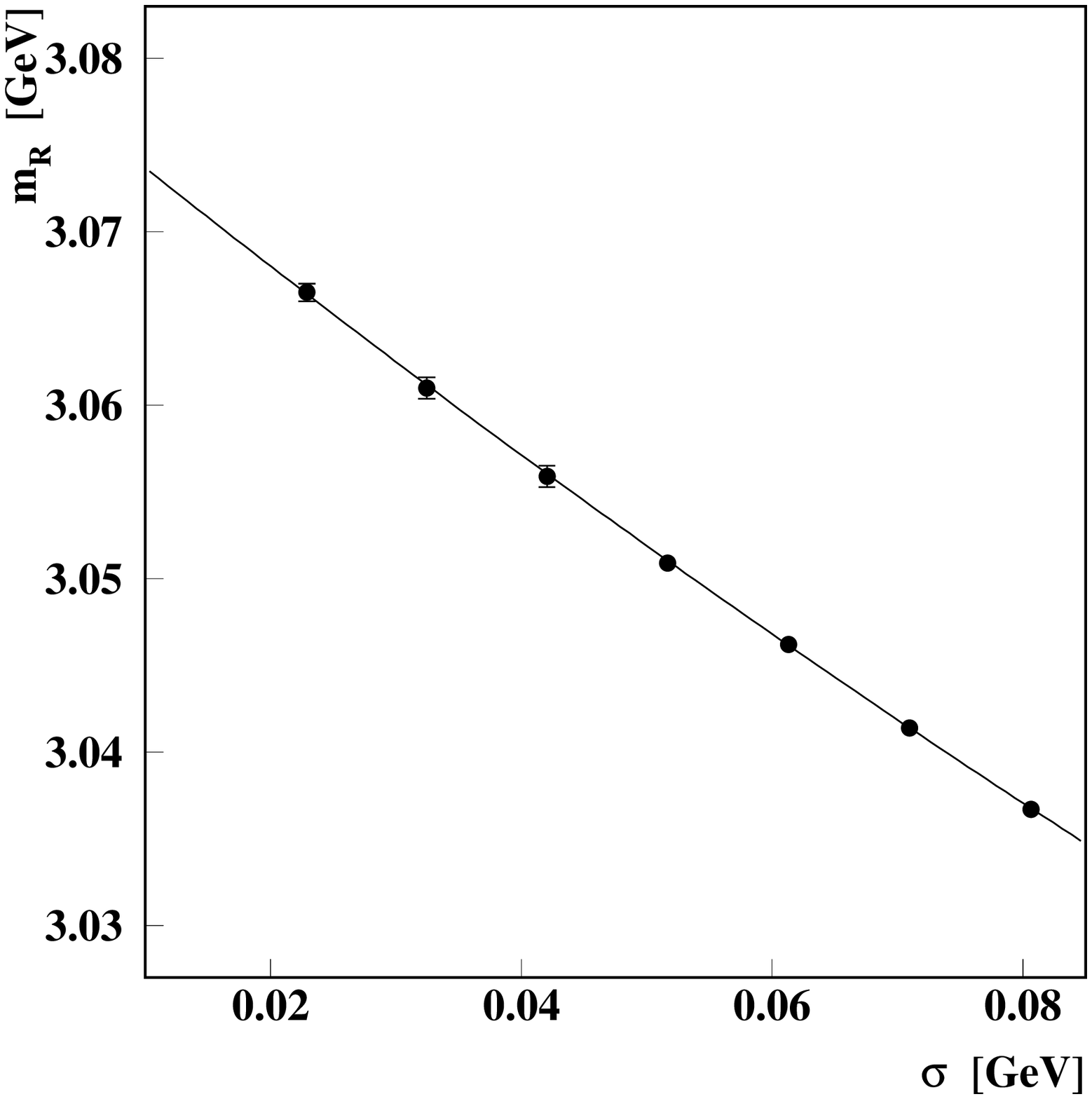,width=8.9cm}}
\put(15.0,4.2){\makebox(0,0)[t]{\huge b)}}
\end{picture}
\caption{\small 
The parameters $\sigma_R$~~(a) and $m_R$~~(b) 
for the
dimuon spectrum from the decay $J/\psi\rightarrow\mu^+\mu^-\gamma$
as functions of
the mass resolution $\sigma$.} 
\label{srmrjpsi}
\end{center}
\end{figure}

The remaining parameters $m_R, \sigma_R, a_i$ were used as free parameters
and determined from the condition of the best fit of  ${\cal P}(m)$
to the dilepton mass distribution for $10^6$ simulated
radiative decays. 
For the decay $J/\psi\rightarrow \mu^+\mu^-\gamma$ 
the parameters $a_1, a_3, a_5, a_{10}$ were found to be statistically 
significant. The function ${\cal P}(m)$ with 
these parameters describe the simulated 
dimuon spectrum quite well, as seen in fig.~\ref{radtailjpsi}.

We repeat the fitting of spectra simulated with different 
detector momentum resolutions
and evaluated the dependences of the considered parameters  
as functions of the mass resolution $\sigma = \sigma(m)$ for $m = M$.
Linear or quadratic polynomials in $\sigma$ were sufficient to
define these functions.   
The parameter $\sigma$ is determined from the
Gaussian fit of the  $\mu^+\mu^-$ mass spectra for independently simulated 
$J/\psi\rightarrow \mu^+\mu^-$ decays. 
\begin{figure}[ht]
\begin{center}
\unitlength1cm
\begin{picture}(15.5,8)
\put( -.50,0.0){\epsfig
{file=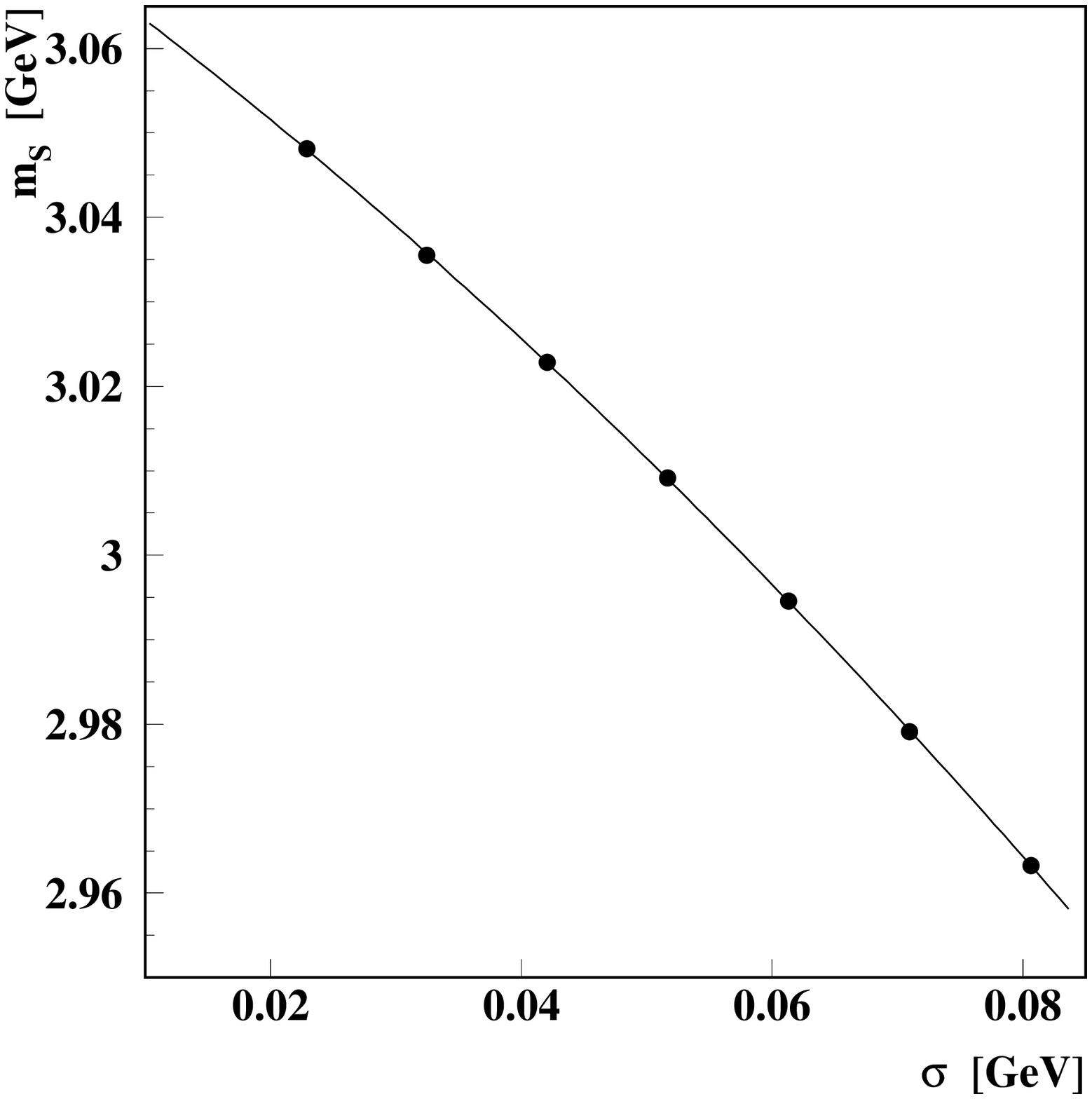,width=8.9cm}}
\put(7.0,4.2){\makebox(0,0)[t]{\huge a)}}
\put(7.7,0.0){\epsfig
{file=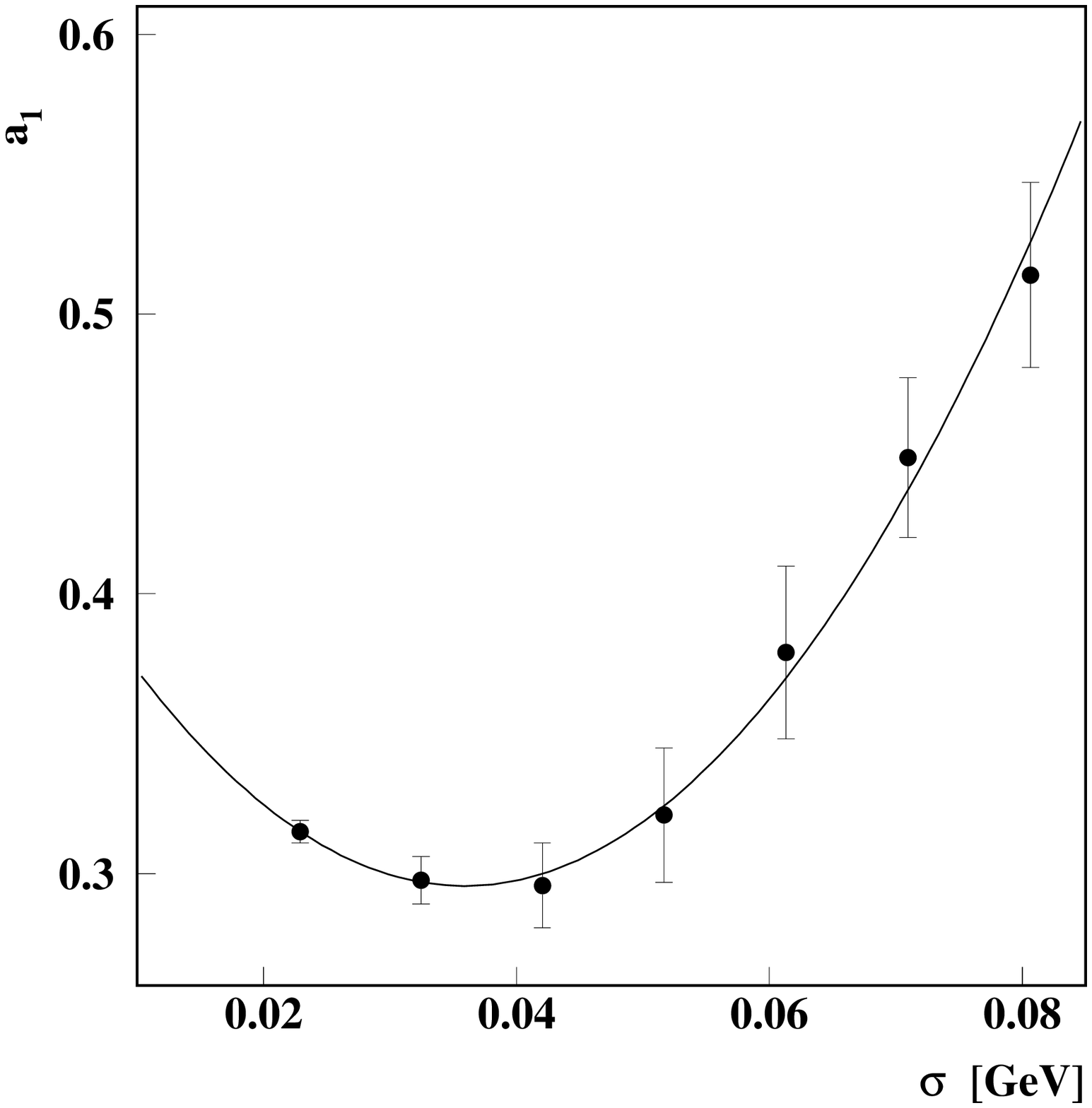,width=8.9cm}}
\put(15.0,4.2){\makebox(0,0)[t]{\huge b)}}
\end{picture}
\caption{\small 
The parameters $m_S$\,~(a) and $a_1$\,~(b) 
for the dimuon spectrum from the decay $J/\psi\rightarrow\mu^+\mu^-\gamma$
as functions of
the mass resolution $\sigma$.} 
\label{switch}
\end{center}
\end{figure}

The parameter $\sigma_R$ depends linearly on the mass resolution
as shown in fig.~\ref{srmrjpsi}a). The nonlinearity of 
the parameter $m_R$ is quite small as seen in fig.~\ref{srmrjpsi}b). 
The point $m_S$, where the Gaussian distribution $G(m, m_R, \sigma_R)$
turns to the long tail, moves towards lower masses for a worse mass
resolution (see fig.~\ref{switch}a). The coefficients
of the polynomial (\ref{sigmas}) were rather well approximated
by quadratic functions as presented in fig.~\ref{switch}b) for
the parameter $a_1$. 

Finally, we approximated as a quadratic polynomial
in $\sigma$ the factor       
\begin{equation}  
N_R (\sigma) = \frac {1} {\int {\cal P}(m|\sigma)\,d\,m}
\label{nr}  
\end{equation}  
to be used for normalisation of the function 
${\cal P}(m|\sigma)$ for the given mass resolution.

As an example, we present the functional dependence for parameters
determined for $J/\psi\rightarrow \mu^+\mu^-\gamma$ decays:  
\[ \sigma_R(\sigma) = 0.22852 \cdot 10^{-2} + 1.0928 \cdot \sigma, \]
\[ m_R(\sigma) = 3.0795 - 0.58797 \cdot \sigma + 0.71851 \cdot \sigma^2, \]
\[ m_S (\sigma) = 3.0744 - 1.0624 \cdot \sigma - 3.9207 \cdot \sigma^2,\]
\[ a_1 (\sigma) = 0.44397 - 8.2646 \cdot \sigma + 115.10 \cdot \sigma^2,\]
\[ a_3 (\sigma) = 2.5218 - 14.572 \cdot \sigma + 479.51 \cdot \sigma^2, \]
\[ a_5 (\sigma) = 3.3055 - 50.711 \cdot \sigma + 636.53 \cdot \sigma^2, \]
\[ a_{10}(\sigma) = 1.4641 - 9.2449 \cdot \sigma + 237.66 \cdot \sigma^2, \]
\[ N_R (\sigma) = 0.34397 + 7.2525 \cdot \sigma - 32.474 \cdot \sigma^2,\]
where $\sigma$ is varying from 0.023 to 0.081 GeV.

\begin{figure}[ht]
\begin{center}
\epsfxsize=0.5\textwidth
\epsfbox{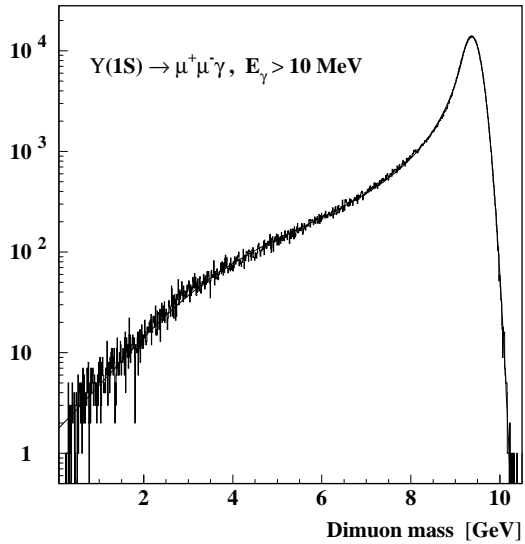}
\caption{\small
Monte Carlo results for the dimuon mass spectrum from the radiative decay
$\Upsilon(1S) \rightarrow \mu^+\mu^-\gamma$ (histogram) and their
parameterisation (solid line). The simulation was made 
for the mass resolution of about 176 MeV
at the $\Upsilon$ peak.}
\label{radtailupsilon}
\end{center}
\end{figure}
The parameterisation describes also well the dimuon mass spectra
simulated for the radiative decay $\Upsilon\rightarrow \mu^+\mu^-\gamma$
as seen in fig.~\ref{radtailupsilon}. 
The shape of the tail looks more complicated than for the $J/\psi$.
The shoulder around 4 GeV can be explained by acceptance effects for
the simulated detector geometry. 
These effects 
appear in the mass range corresponding to very low values
of ${\cal P}(m|\sigma)$.    
An additional parameter
$a_{20}$ was needed in polynomial (\ref{sigmas}) to overcome the 
complications of the mass spectrum.

\begin{table}[ht]
\begin{center}
\caption{\small 
Sets of coefficients of the polynomial (\ref{sigmas}) 
found to be sufficient to parameterise 
inclusive dilepton mass spectra simulated for
radiative decays of different particles.}
\vspace*{5mm}
\label{aall}
\begin{tabular}{|c|c|c|}
\hline \hline
  Particle  & Decay into $\mu^+\mu^-\gamma$ & Decay into $e^+e^-\gamma$  \\
\hline 
\hline
$J/\psi$ & $a_1, a_3, a_5, a_{10}$ & $a_1, a_3, a_5, a_{10}, a_{20}$ \\
\hline
 $\psi(2S)$ & $a_1, a_3, a_5, a_{10}$ & $a_1, a_3, a_5, a_{10}, a_{20}$ \\
\hline
$\Upsilon(1S)$&$a_1, a_3, a_5, a_{10}, a_{20}$ &$a_1, a_3, a_5, a_{10}, a_{28}$ \\
\hline
\hline
\end{tabular}
\end{center}
\end{table}
We simulated radiative decays of the $J/\psi, \psi(2S), \Upsilon(1S)$
into muon or electron pairs 
and evaluated the corresponding parameterisations for the dilepton spectra. 
In all cases, the parameters $a_1, a_3, a_5, a_{10}$ were found to be
significant as shown in table~\ref{aall}. For decays with
smaller $m_l / M$  an additional parameter was needed
($a_{20}$ or $a_{28}$), to take into account specific details induced by   
the simulated experimental set-up. These parameters are needed 
at the lower edge
of the mass spectra where the probability densities are low.

 \section{Treatment of Inclusive Measurements of Dilepton Mass Spectra}
The dilepton mass spectra measured inclusively
can be described as 
\begin{equation}
S(m\,|\,M, \sigma) = 
(1 - W_{hard}) \cdot G(m | M, \sigma) 
\,\,+\,\,W_{hard} \cdot N_R(\sigma) \cdot {\cal P}(m |\,M, \sigma), 
\label{pplusg}  
\end{equation}  
where ${\cal P}(m|\,M, \sigma)$ is the function (\ref{gauss},\ref{gtail})
with corresponding parameterisation normalised by the factor $N_R(\sigma)$
and the weight $W_{hard}$ defines 
the relative contribution of the radiative tail into 
the spectrum. The first term
describes the configurations without photon  
and with low energies photons emissions 
smeared accordingly by detector resolutions.
In a real experiment this term could be a more complicated
function than a Gaussian. 
For electrons also an additional term
could be appropriate to describe  
the bremsstrahlung process on the path through the detector
material. As it was mentioned already, this process was not
simulated for the present study.
The weight $W_{hard}$ can be defined as
\begin{equation} 
W_{hard} = \frac {C_{hard}(E_{min}) \cdot A_{hard}(E_{min})}
{(1 + C_{soft}(E_{min})) \cdot A_{soft}(E_{min})
+ C_{hard}(E_{min}) \cdot A_{hard}(E_{min})},
\label{wharda}   
\end{equation}   
where $A_{soft}(E_{min})$ is an acceptance for $l^+l^-$ and 
$l^+l^-\gamma,\,(E_{\gamma} \le E_{min})$
final states and
$ A_{hard}(E_{min})$ that for $l^+l^-\gamma,\,(E_{\gamma} > E_{min}).$ 
For close $A_{soft}$ and $A_{hard}$
the definition of $W_{hard}$ by eq.~(\ref{whard}) 
can be sufficient.
The latter definition for $E_{min} = 10$ MeV we use for further calculations.
\begin{figure}[ht]
\begin{center}
\epsfxsize=0.55\textwidth
\epsfbox{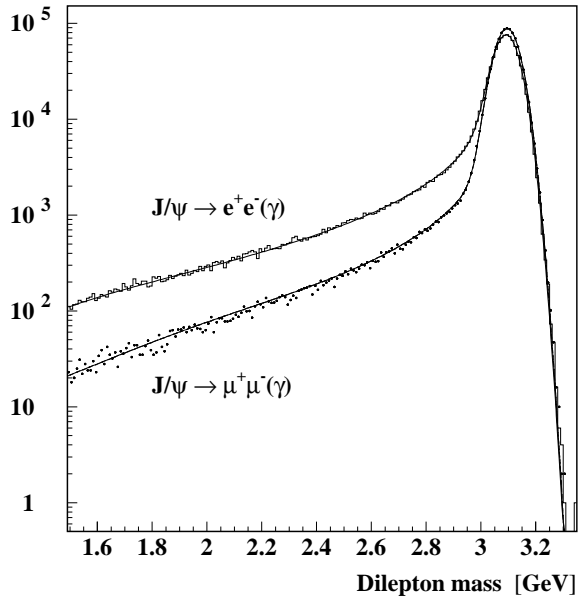}
\caption{\small
Monte Carlo results for inclusive dilepton mass spectra from 
decays of $J/\psi$ into $e^+e^-$, $e^+e^-\gamma$ (histogram)
and $\mu^+\mu^-$, $\mu^+\mu^-\gamma$ (points)
and the fit curves using the 
parameterisations (\ref{pplusg})(solid curves). The mass resolution 
was about 42 MeV at the $J/\psi$ peak for both decays.}
\label{jpsiall}
\end{center}
\end{figure}

The corresponding mass spectra for the $J/\psi$ decays are presented
in fig.~\ref{jpsiall}. The integral of the function (\ref{pplusg})
over the full interval of dilepton mass equals unity by definition.
For the fitting of the histogram we used 
the product of the corresponding function (\ref{pplusg}) 
and the number of events treated as fit parameter.
The function (\ref{pplusg}) itself depends on two parameters:
the mass average $M$ and resolution $\sigma$ of the Gaussian part.
Hence, there are three fit parameters in total.
 
\begin{figure}[ht]
\begin{center}
\unitlength1cm
\begin{picture}(15.5,8.5)
\put( -.50,0.0){\epsfig
{file=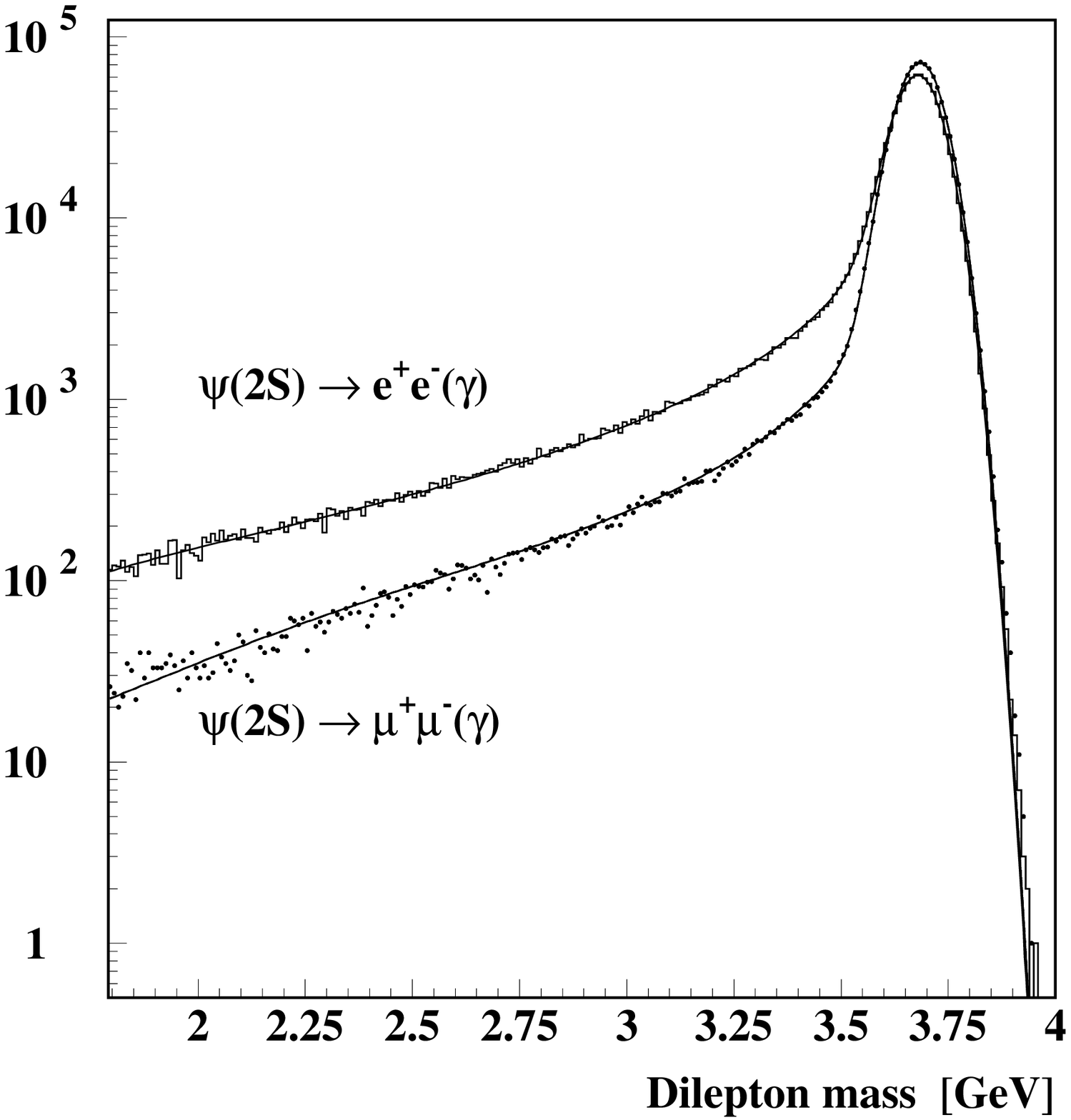,width=8.9cm}}
\put(1.5,7.6){\makebox(0,0)[t]{\huge a)}}
\put(7.9,0.0){\epsfig
{file=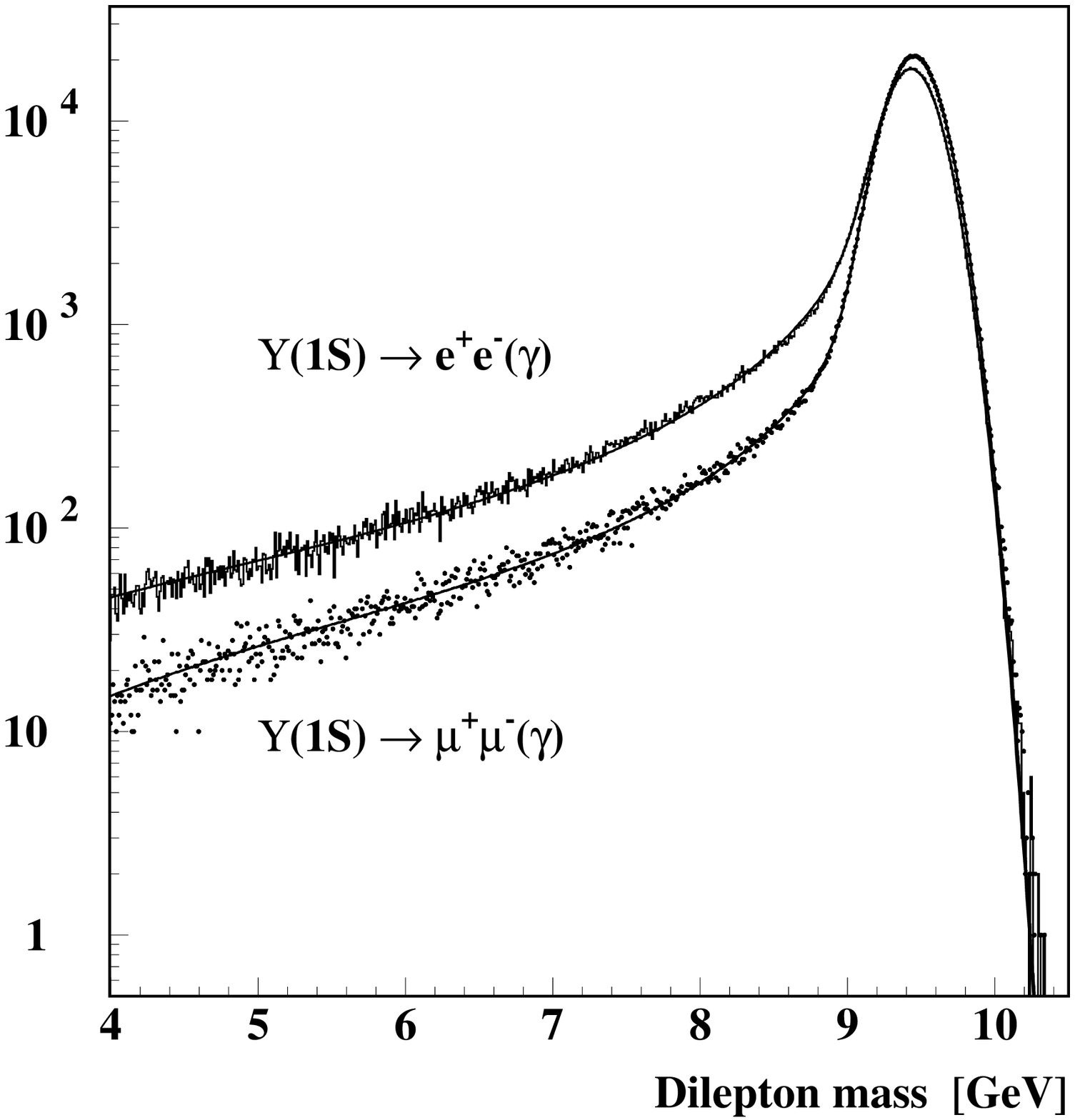,width=8.9cm}}
\put(10.,7.6){\makebox(0,0)[t]{\huge b)}}
\end{picture}
\caption{\small 
Monte Carlo results for inclusive dilepton mass spectra from 
onia decays into $e^+e^-$, $e^+e^-\gamma$ (histogram)
and $\mu^+\mu^-$, $\mu^+\mu^-\gamma$ (points)
and  parameterisations (\ref{pplusg})(solid curves). 
a)~Decays of the $\psi(2S)$ were simulated with a mass
resolution of 52 MeV at the signal peak.
b)~Decays of the $\Upsilon(1S)$ for a mass
resolution of 176 MeV at the signal peak.}
\label{psi2sups}
\end{center}
\end{figure}

The fitted curves describe well the results of the simulation
as shown in fig.~\ref{jpsiall}.
The fitting procedures for 
decays of the $\psi(2S)$ shown in fig.~\ref{psi2sups}a)
and $\Upsilon(1S)$ in fig.~\ref{psi2sups}b) were analogous
to those described above. The fitting function also agreed
with the simulation in a wide range of invariant masses. 

\begin{figure}[ht]
\begin{center}
\unitlength1cm
\begin{picture}(15.5,6)
\put(7.39,0.0){\epsfig
{file=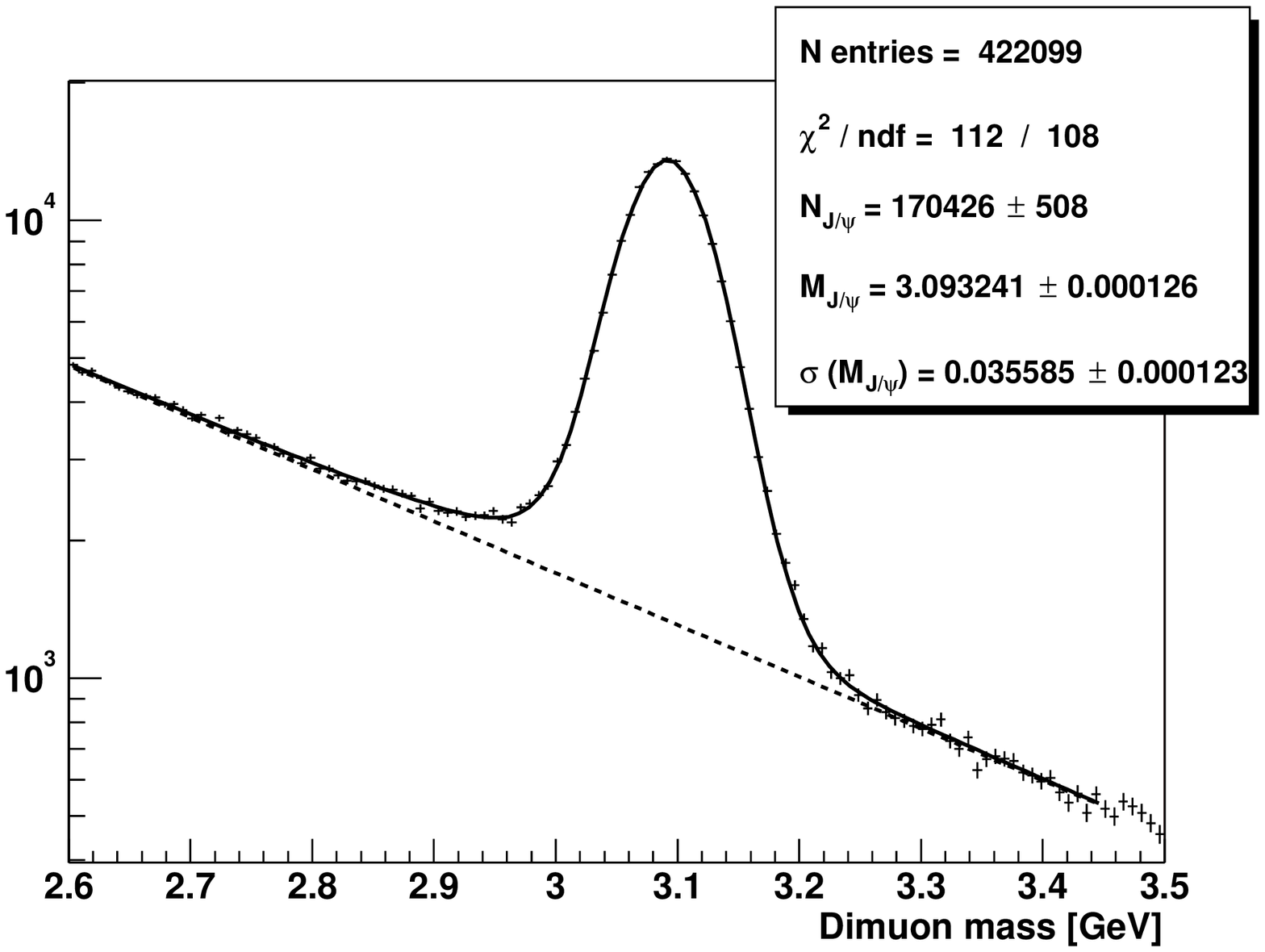,width=8.7cm}}
\put( -1.0,0.0){\epsfig
{file=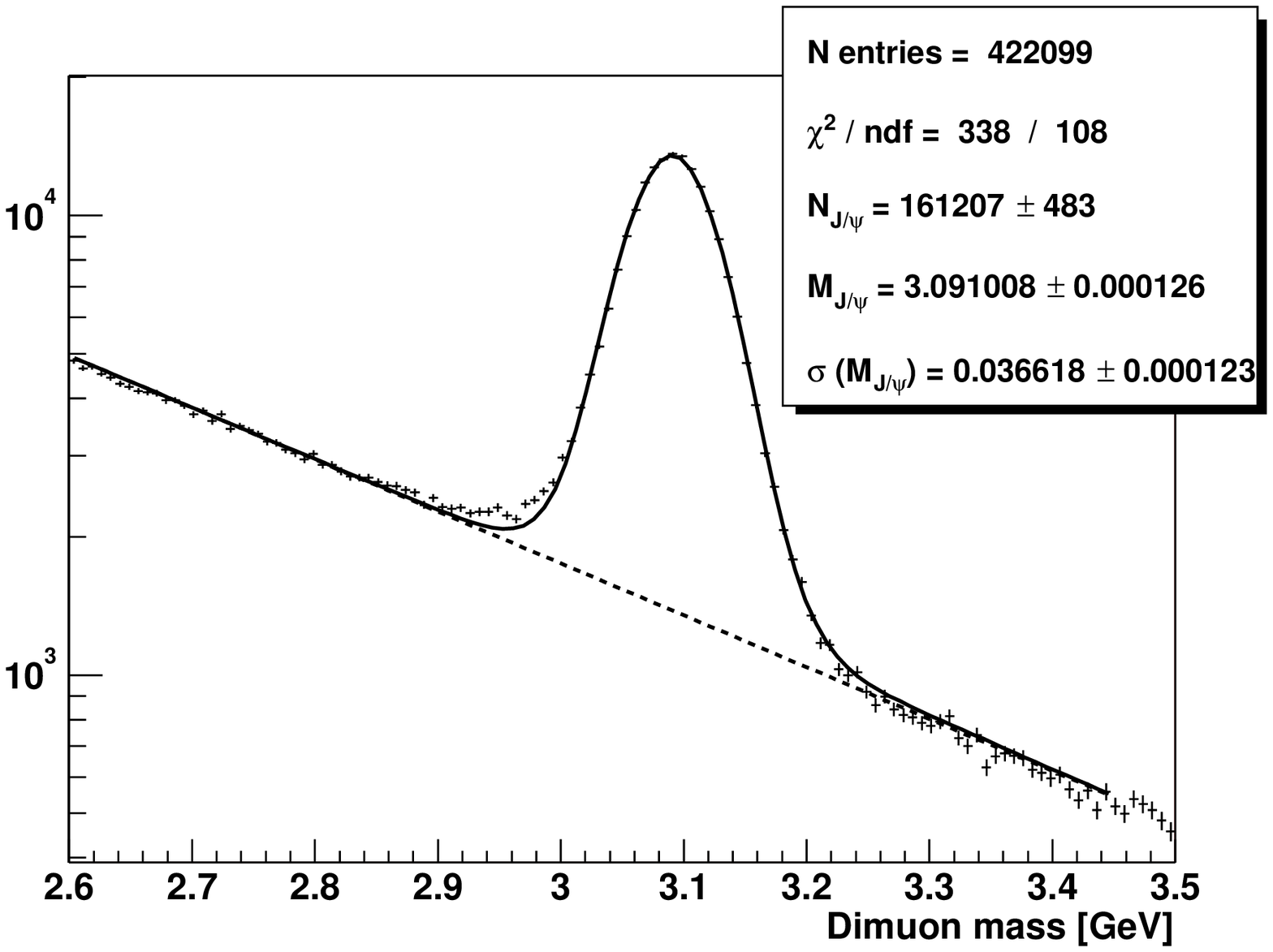,width=8.9cm}}
\put(1.1,2.2){\makebox(0,0)[t]{\huge a)}}
\put(9.6,2.2){\makebox(0,0)[t]{\huge b)}}
\end{picture}
\caption{\small 
The invariant mass distribution of $\mu^+\mu^-$ around the $J/\psi$ peak
measured in the HERA-B experiment and fitted without the radiative tail 
taken into account (a) 
and with the parameterisation including the radiative tail (b).
The values of $\chi^2$ were 338 and 112, respectively, for 108 degrees
of freedom. The background was described as exponential function
(dashed curves). 
In inserts are shown: number of entries in the histogram,\,
$\chi^2$,\,fitted number of $J/\psi$,\,  
fitted mass and mass resolution in GeV.}
\label{real}
\end{center}
\end{figure}

To make a test on real data, we used 
the HERA-B data recorded with a dimuon trigger during the 2002--2003
HERA running period.
Figure~\ref{real} shows the dimuon mass spectrum obtained from these data
with a prominent signal above some background.
The background was described by an exponential function. 

In the HERA-B detector, a forward spectrometer 
with non-negligible amount of material on the path through 
the magnet~\cite{fsres},
Moli\`{e}re scattering is essential~\cite{momres} for the momentum 
resolution.  
Therefore in the fitting function (\ref{pplusg}) instead of one 
Gaussian we used a symmetric function composed of three Gaussians
to take into account the details of the momentum (mass) resolution 
in the HERA-B detector. 
The fit of the dimuon mass spectrum
by this symmetric function accompanied by an exponential 
function for the background
resulted in $\chi^2$ = 338 for 108 degrees of freedom
as presented in ~fig.\ref{real}a).
\clearpage
\clearpage 
The inclusion of the radiative tail 
into the fitting function improved the result as seen in ~fig.\ref{real}b),
where $\chi^2$ = 112 for 108 degrees of freedom. 
One should remember that the radiative tail
(second term in (\ref{pplusg}))
was defined through parameters of the first term,
i.e. $M$ and $\sigma$ of the main peak, and the number of fitted parameters,
and thus the degrees of freedom, were the same. 
Taking the radiative tail into account  in the inclusive spectra
of the dilepton mass could be significant for large statistics and 
rather high mass resolution.  
\section{Summary}
Well known analytic results for the radiative decay
of the vector boson $Z\rightarrow l^+l^-\gamma$ valid for arbitrary 
cuts in the photon energy and angles were used to estimate contributions
of similar decays for charmonia and bottomonia vector states.
The contributions 
were found to be rather large not only for decays
into electrons but also into muons. 
The radiative effects appeared as tails
in inclusive mass spectra of the lepton pairs.
An analytic formula for the dilepton mass spectra in 
radiative decays is derived. The spectra convoluted 
with the detector mass resolution could
be described by rather simple parameterisations.
The parameterisations were studied by 
Monte Carlo for $J/\psi, \psi(2S), \Upsilon(1S)$
radiatively  decaying into $\mu^+\mu^-\gamma$ or $e^+e^-\gamma$
using a simulation describing the real experiment.
The parameterisation described the Monte Carlo results quite
well and was found to be appropriate also for fitting real data.        
\vspace{6mm}
\newline
{\bf Acknowledgments:}
I had great benefit from the expertise and advices of F.\,Jegerlehner
whose detailed analytic results for radiative decays of Z bosons were
the starting point for this study. 
Cordial thanks to T.\,Lohse who stressed the importance of
radiative decays for charmonia measurements and developed 
the Monte Carlo generator used for the investigation.
Fruitful discussions with G.\,Bohm and H.\,Kolanoski helped me
a lot to finalize the results. I am also grateful to them
for the careful reading of the manuscript.
I would like to thank DESY Zeuthen for the kind hospitality extended 
to me during my visit.
%
%

\end{document}